\documentclass[conference]{IEEEtran}
\IEEEoverridecommandlockouts
\usepackage{cite}
\usepackage{amsmath,amssymb,amsfonts}

\usepackage{graphicx}
\usepackage{textcomp}
\usepackage{xcolor}
\usepackage{url}

\usepackage{algorithm}
\usepackage[noend]{algpseudocode}
\algnewcommand\algorithmicforeach{\textbf{for each}}

\newcommand{\REGISTERED}{\textsuperscript{\textregistered}}

\algdef{S}[FOR]{ForEach}[1]{\algorithmicforeach\ #1\ \algorithmicdo}

\def\BibTeX{{\rm B\kern-.05em{\sc i\kern-.025em b}\kern-.08em
    T\kern-.1667em\lower.7ex\hbox{E}\kern-.125emX}}

\definecolor{Green}{rgb}{0, 0.5, 0}
\definecolor{cust-green}{rgb}{0.0, 0.5, 0.0}

\usepackage{pgfplots}  
\pgfplotsset{width=6.6cm,compat=1.7}  
\usepackage{tikz}

\definecolor{bblue}{HTML}{4F81BD}
\definecolor{rred}{HTML}{C0504D}
\definecolor{ggreen}{HTML}{9BBB59}
\definecolor{ppurple}{HTML}{9F4C7C}
\definecolor{aureolin}{rgb}{0.99, 0.93, 0.0}
\definecolor{brown(traditional)}{rgb}{0.59, 0.29, 0.0}
\definecolor{grey}{rgb}{0.7, 0.75, 0.71}

\definecolor{ColorKD}{rgb}{0.1,0.6,0.3}
\newboolean{MARKUP}
\setboolean{MARKUP}{true}

\newcommand{\tsSAY}[1]{\slshape\sffamily\color{#1}}
\newcommand{\IFXSAY}[3]{{\ifthenelse{\boolean{MARKUP}}{\tsSAY{#2} #1:~#3}{}}}

\begin{document}
\title{MetaFI: Model-driven Fault Simulation Framework}

\author{%
	\IEEEauthorblockN{%
		Endri Kaja\IEEEauthorrefmark{1}\IEEEauthorrefmark{2},  
		Nicolas Gerlin\IEEEauthorrefmark{1}\IEEEauthorrefmark{2},
		Luis Rivas\IEEEauthorrefmark{1}\IEEEauthorrefmark{2},\
		Bora Monideep\IEEEauthorrefmark{1}\IEEEauthorrefmark{4},\\
		Keerthikumara Devarajegowda\IEEEauthorrefmark{1}\IEEEauthorrefmark{2},
		Wolfgang Ecker\IEEEauthorrefmark{1}\IEEEauthorrefmark{3}
	}
	\IEEEauthorblockA{%
				\IEEEauthorrefmark{1}Infineon Technologies AG, Germany\\
				\IEEEauthorrefmark{2}Technische Universit\"at Kaiserslautern, Germany\\
				\IEEEauthorrefmark{4}Albert-Ludwigs-Universit\"at Freiburg, Germany\\
				\IEEEauthorrefmark{3}Technische Universit\"at M\"unchen, Germany
	}
}
\maketitle
\begin{abstract}

Safety-critical designs need to ensure reliable operations under hostile conditions with a certain degree of confidence. The continuously higher complexity of these designs makes them more susceptible to the risk of failure. ISO26262 recommends fault injection as the proper technique to verify and measure the dependability of safety-critical designs. To cope with the complexity, a lot of effort and stringent verification flow is needed. Moreover, many fault injection tools offer only a limited degree of controllability.

We propose MetaFI, a model-driven simulator-independent fault simulation framework that provides multi-purpose fault injection strategies such as Statistical Fault Injection, Direct Fault Injection, Exhaustive Fault Injection, and at the same time reduces manual efforts. The framework enables injection of Stuck-at faults, Single-Event Transient faults, Single-Event Upset faults as well as Timing faults. The fault simulation is performed at the Register Transfer Level (RTL) of a design, in which parts of the design targeted for fault simulation are represented with Gate-level (GL) granularity. MetaFI is scalable with a full System-on-Chip (SoC) design and to demonstrate the applicability of the framework, fault simulation was applied to various components of two different SoCs. One SoC is running the Dhrystone application and the other one is running a Fingerprint calculation application. A minimal effort of 2 person-days was required to run 38 various fault injection campaigns on both the designs. The framework provided significant data regarding failure rates of the components. Results concluded that Prefetcher, a component of the SoC processor, is more susceptible to failures than the other targeted components on both the SoCs, regardless of the running application.

\end{abstract}


\section{Introduction}
\label{sec:Introduction}

Semiconductor devices operate in various environmental conditions for a long duration of time.
Natural radiation, aging, mechanical stress, processing defects, and other events impact the intended behavior of the system by introducing numerous faults. When a fault causes an incorrect behavior of the device, a failure occurs. Due to the increased complexity and growing transistor density of modern designs, the probability of failures is noticeably becoming higher \cite{Oetjens}. Therefore, fault detection and correction mechanisms are integrated into safety-critical designs to detect failure events and take necessary precautionary measures. ISO26262 \cite{ISO26262}, the functional safety standard for the automotive domain, recommends \textit{fault injection} as the proper technique to assess circuit and system dependability. 

Fault injection is defined by \cite{arlat}, \cite{ziade} as the validation technique of fault tolerant systems by observing the behavior of the system in the presence of injected faults. Fault injection is a well-researched topic and most of the developed techniques fall into five categories\cite{ziade}:
\begin{itemize}
    \item Hardware-based fault injection,
    \item Software-based fault injection,
    \item Simulation-based fault injection,
    \item Emulation-based fault injection,
    \item Hybrid fault injection.
\end{itemize}

In this paper, we propose MetaFI, a fully automated model-driven simulation-based fault injection framework that reduces manual efforts significantly. Initially, an existing saboteur-based fault injection flow of mixed RTL/GL granularity is extended to support Timing faults and Single-Event Upsets. Next, a \textit{metamodel}-based generator framework is utilized to automatically support various fault injection strategies such as Exhaustive Fault Injection, Statistical Fault Injection, Direct Fault Injection, and at the same time to provide a simulator-independent fault simulation campaign.

To demonstrate the applicability, scalability, and advantages of the proposed approach, various fault simulation campaigns are applied to different target components of two distinct RISC-V based SoCs, running a \textit{Fingerprint calculation application} and \textit{Dhrystone} program respectively. Experimental results show that only a minimal amount of manual effort of 2 man-days was required to run 38 different fault injection campaigns. Further, it was observed that for every campaign we ran, the Prefetcher component, responsible for program flow, had the highest rate of failures on both the SoCs and we recommended integrating different protection mechanisms to this component to reduce the risk of failures.

The paper is structured as follows: Section II describes the existing fault injection flow. In Section III, we present the extended fault models to support Timing faults and SEUs. Section IV gives a detailed description of the model-based fault simulation framework. The experimental results are described in Section V and a brief summary in Section VI 
completes the paper.

\section{Background}

\subsection{Fault models}
Fault models are used to abstract the effect of faults onto different abstraction levels of digital designs. The widely applied fault models on the circuit and transistor-level of the design are \textit{permanent} and \textit{transient} faults. 

Permanent faults (hard errors) remain in the system for the full duration of the operation, or until they are externally fixed. These kinds of faults are mainly caused by physical effects, e.g. processing defects,  electromigration, and other aging effects \cite{permanent_fault}.  Some examples of permanent faults are \textit{stuck-at-0}, i.e. a signal is permanently tied to logical value 0, \textit{Stuck-at-1}, i.e. a signal is permanently tied to logical value 1, \textit{Timing faults}, i.e. a delayed response in the system.

Transient faults (soft errors) are introduced in the system only for a short duration of time and are mainly caused by radiation effects \cite{transient_fault}. These kinds of faults create a temporary change in the voltage of a signal that causes a logical bitflip. Common examples of transient faults are \textit{Single-Event Transients}, i.e. a bitflip in the output of a combinational gate, \textit{Single-Event Upset}, i.e. a reversible bitflip in a sequential cell (memory, registers, latches).

\subsection{Fault Injection Flow}
\label{sec:Fault injection flow}

Fault modeling enables analyzing the effects of the faults on the design utilizing fault injection. Traditionally, fault injection has been applied to the Gate-Level (GL) and Register Transfer Level (RTL) granularity of the design. Although fault injection at GL granularity provides accurate \textit{fault coverage} \cite{Kochte}, \cite{syfi}, it suffers from low performance. Contrary, RTL fault injection increases the performance but limits the applicable fault models.

Ecker et al. \cite{kaja} presented a mixed RTL/GL fault injection flow that  utilizes an existing metamodel-based RTL generation framework \cite{Schreiner.etal-2016},  \cite{Schreiner.etal-2017} to generate the RTL on a mixed granularity that enables injecting different categories of faults.
 \begin{figure}[!tbh]
            	\begin{center}
            		\includegraphics[width=\linewidth]{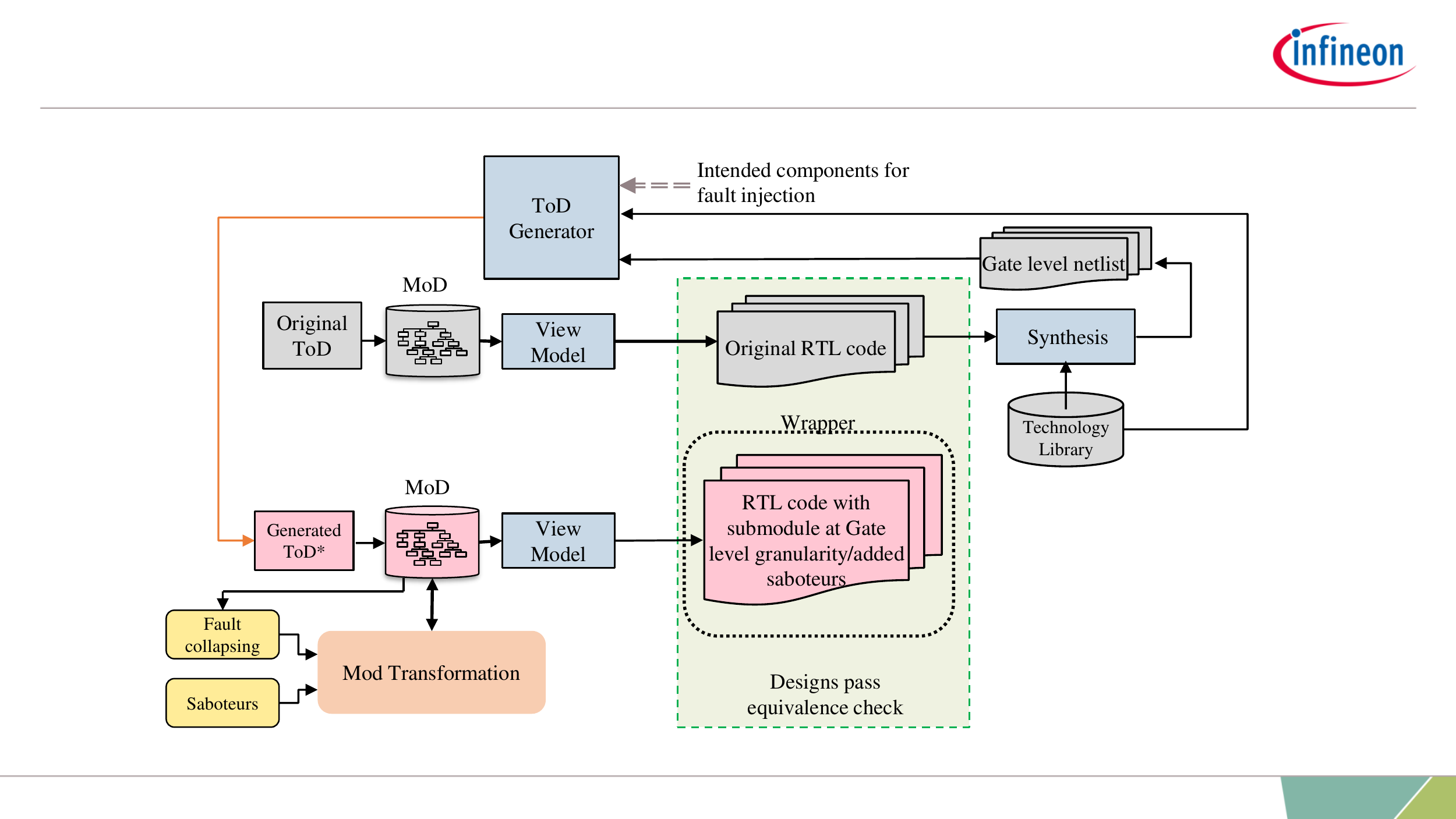}
            		\vspace{-5mm}
            	\end{center}
            	\caption{Mixed register transfer/gate-level design generation flow}
            	\label{fig:fault_injection_flow}
            	\vspace{-3mm}
            \end{figure}
            
RTL generation is composed of three main abstraction levels. The first level captures the specifications into formal models through a metamodel. The specification models are named as Model-of-Things (MoTs), and after the creation of the MoT, a Python-based framework utilizes the microarchitecture blueprint defined in the Template-of-Desing (ToD) to generate a platform and technology independent hardware description level termed as Model-of-Design (MoD). Lastly, the MoD is mapped to Model-of-View (MoV) to generate the RTL in a selected Hardware Description Language (HDL). 

The aforementioned flow is extended to generate the RTL on a mixed RTL/GL granularity as shown in Fig. \ref{fig:fault_injection_flow}. Initially, the generated RTL is synthesized to get its Gate-level representation, generally known as GL netlist. The ToD Generator (Fig. \ref{fig:fault_injection_flow}) extracts the data from the netlist and the specific technology library used to synthesize the design. Depending on the input from the user, i.e. component that is subject to fault injection, ToD Generator generates a new ToD with a mixed RTL/GL granularity. For example, let us consider a CPU core and the Arithmetic-Logic Unit (ALU) is the target to apply fault injection. The generator maps the GL representation of the ALU to MoD components and the rest of the design modules are mapped to the original RTL MoD components. Therefore, the ToD generator builds the complete design model with partial RTL and partial GL granularity.

MoD, a platform-independent layer, is a tree-like data structure that can be transformed by adding, deleting, or modifying its items. Consequently, it permits adding extra components, termed \textit{saboteurs} into the design connections to enable fault injection. Saboteurs are particular hardware modules added to intended fault injection locations. They are easily controllable and, when activated, modify the original value of a signal \cite{FI-techniques}. Before adding the saboteurs to the design MoD, \textit{fault collapsing} \cite{F-collapsing} is performed to reduce the total number of injectable faults. The collapsed fault list provides only one of the two stuck-at fault models for equivalent faults (e.g. a stuck-at-0 at the input and output of an AND gate). Saboteur control signals are propagated as primary input ports of the design, allowing a high degree of controllability.

Lastly, an equivalence check is applied between the original RTL design and the transformed mixed RTL/GL granularity design by constraining the saboteurs to not inject any fault.

\section{Extending the fault injection flow }
\label{sec:Extending_the_fault_injection_flow}

The presented flow is insufficient to perform fault injection considering all circuit-level fault models. We propose techniques to inject Timing faults via RTL primitives and extend the current flow to enable SEU independently from the register/latch type. 

\subsection{Timing Faults}

Timing faults, also known as transient-delay faults, were first mentioned by Breuer \cite{Breuer}. These faults arise as a result of manufacturing \cite{Luong}, \cite{Malaiya} and manifest as a delayed response in the system. The affected system parts calculate the correct result, but the response is observed later than in the fault-free scenario. 

 \begin{figure}[!tbh]
                \vspace{-4mm}
            	\begin{center}
            		\includegraphics[scale=0.5]{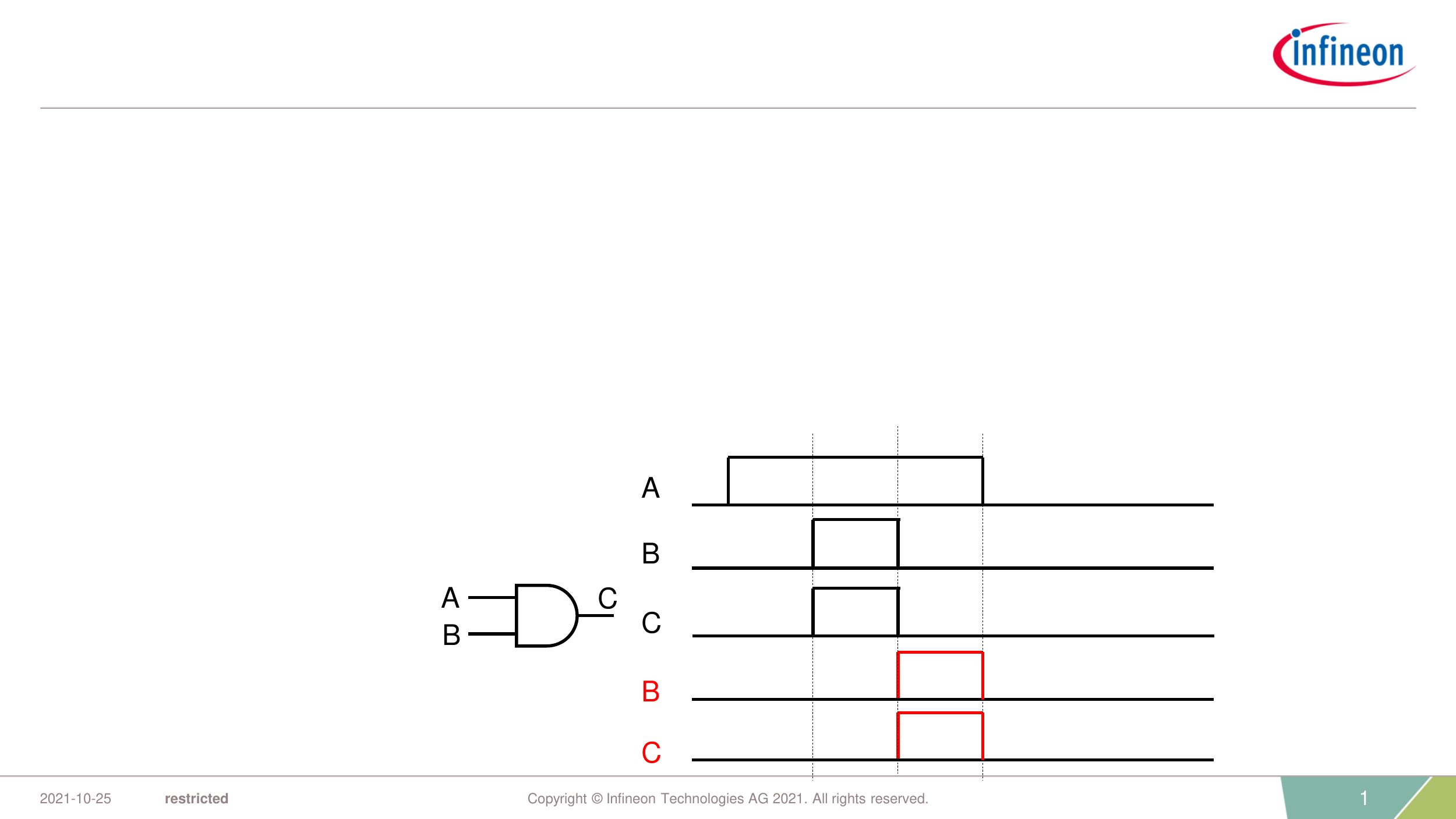}
            		\vspace{-5mm}
            	\end{center}
            	\caption{Timing fault example waveform}
            	\label{fig:tf_waveform}
            	\vspace{-4mm}
            \end{figure}

Assume an AND gate presented in the left side of Fig. \ref{fig:tf_waveform}. Input A has a logical value of 1 for three consecutive clock cycles, input B becomes 1 next clock cycle after A becomes 1 and, consequently, C becomes 1 at the same clock cycle. Due to a timing fault at the AND gate input, the rising edge of input B is delayed by one clock cycle and, therefore, the correct value at the output port is delayed as well.

\begin{figure}[!tbh]
            \vspace{-4mm}
            	\begin{center}
            		\includegraphics[scale=0.6]{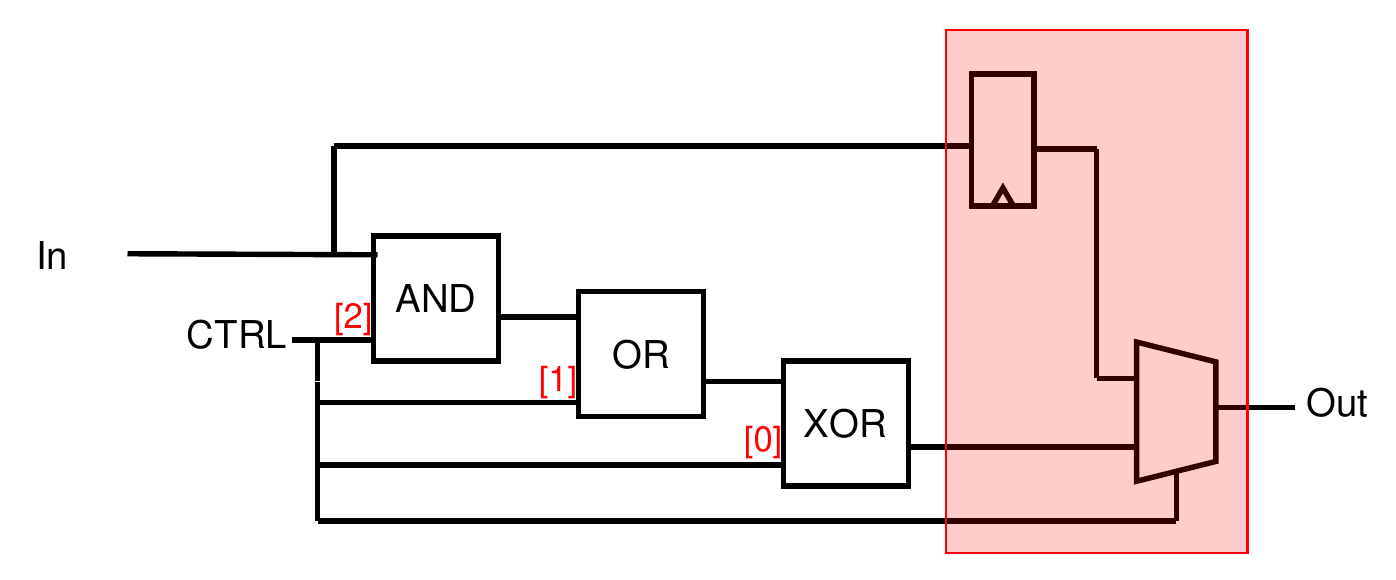}
            		\vspace{-5mm}
            	\end{center}
            	\caption{Timing fault saboteur}
            	\label{fig:timing_fault_saboteur}
            	\vspace{-5mm}
            \end{figure}

To mimic the behavior displayed above, the existing saboteurs in \cite{kaja} are extended as shown in Fig. \ref{fig:timing_fault_saboteur}.
Since the injection flow utilizes a mixed RTL/GL granularity, it is possible to use RTL primitives, i.e. registers to induce a one clock cycle delay. When the last bit of control signal \textit{CTRL} is set to 1, the \textit{In} signal value at the output \textit{Out} is delayed by one clock cycle. Hence, a timing fault is injected to the desired signal location. 

\subsection{Single Event Upset}

Single Event Upset faults occur when radiation causes enough disturbance to flip the logical value of a memory cell, register, latch or flip flop. These soft errors have the ability and potential to cause the highest failure rate of all reliability mechanisms included \cite{Baumann}. The existing saboteurs in the presented fault injection flow trigger a bitflip at the intended signal location. Nevertheless, some state-saving Gate-level technology cells (registers, latches) depend on an enable signal to activate the fault, i.e. if the cell is not activated, the fault is masked. If the testing pattern does not consider the case, it could potentially lead to a dangerous behavior of the design and affect the dependability analysis.

 \begin{figure}[!tbh]
            	\vspace{-4mm}
            	\begin{center}
            		\includegraphics[scale=0.5]{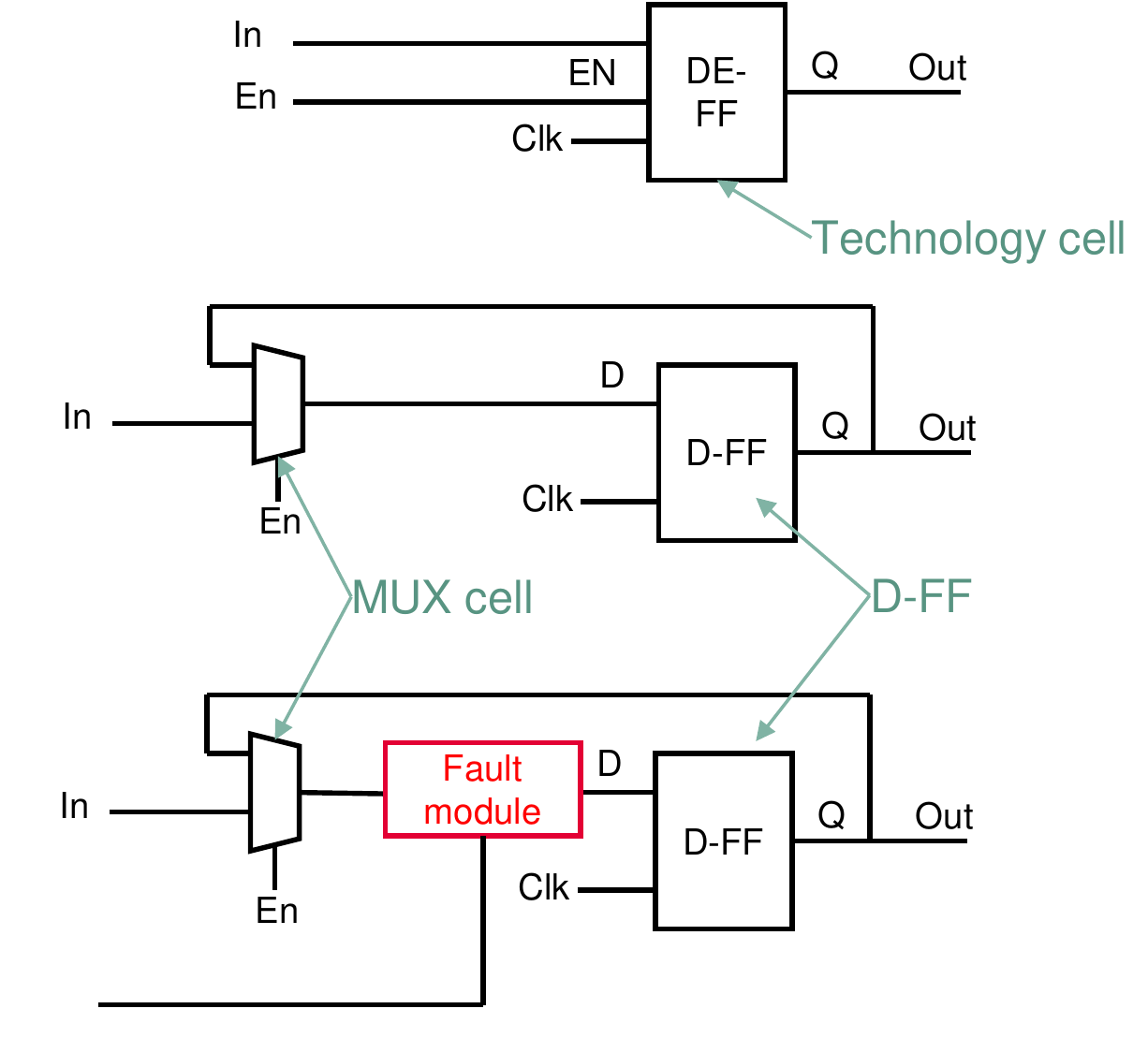}
            		\vspace{-5mm}
            	\end{center}
            	\caption{Injection of SEU into Enabled D-FlipFlop}
            	\label{fig:injection_seu}
            	\vspace{-4mm}
            \end{figure}

To tackle this scenario, we propose to replace enabled technology cells with simple cells that have the enable signal as a select signal for the multiplexer connected to the cell input. An example of transforming an enabled D-Flipflop is presented in Fig. \ref{fig:injection_seu}. Both the circuits are equivalent, i.e. the cell would be written only when the enable signal is activated. Therefore, by adding the saboteurs at the input of the cell, as shown in color red in Fig. \ref{fig:injection_seu}, the value of the cell would be flipped independently of the enable signal.

\section{Model-driven fault simulation framework}
\label{sec:msafety}

The existing fault injection flow enables a high level of controllability of fault injection locations via propagated saboteur control lines as primary inputs. Nevertheless, the process of fault simulation, i.e. determining the fault simulation purpose and selecting fault model, requires manual efforts and is prone to errors. We propose MetaFI, a simulator independent model-driven fault simulation framework that facilitates and automates the process of fault simulation with substantial reduced manual efforts

\subsection{MetaFI Metamodel}

\begin{figure}[!tbh]
            	\begin{center}
            		\includegraphics[width=\linewidth]{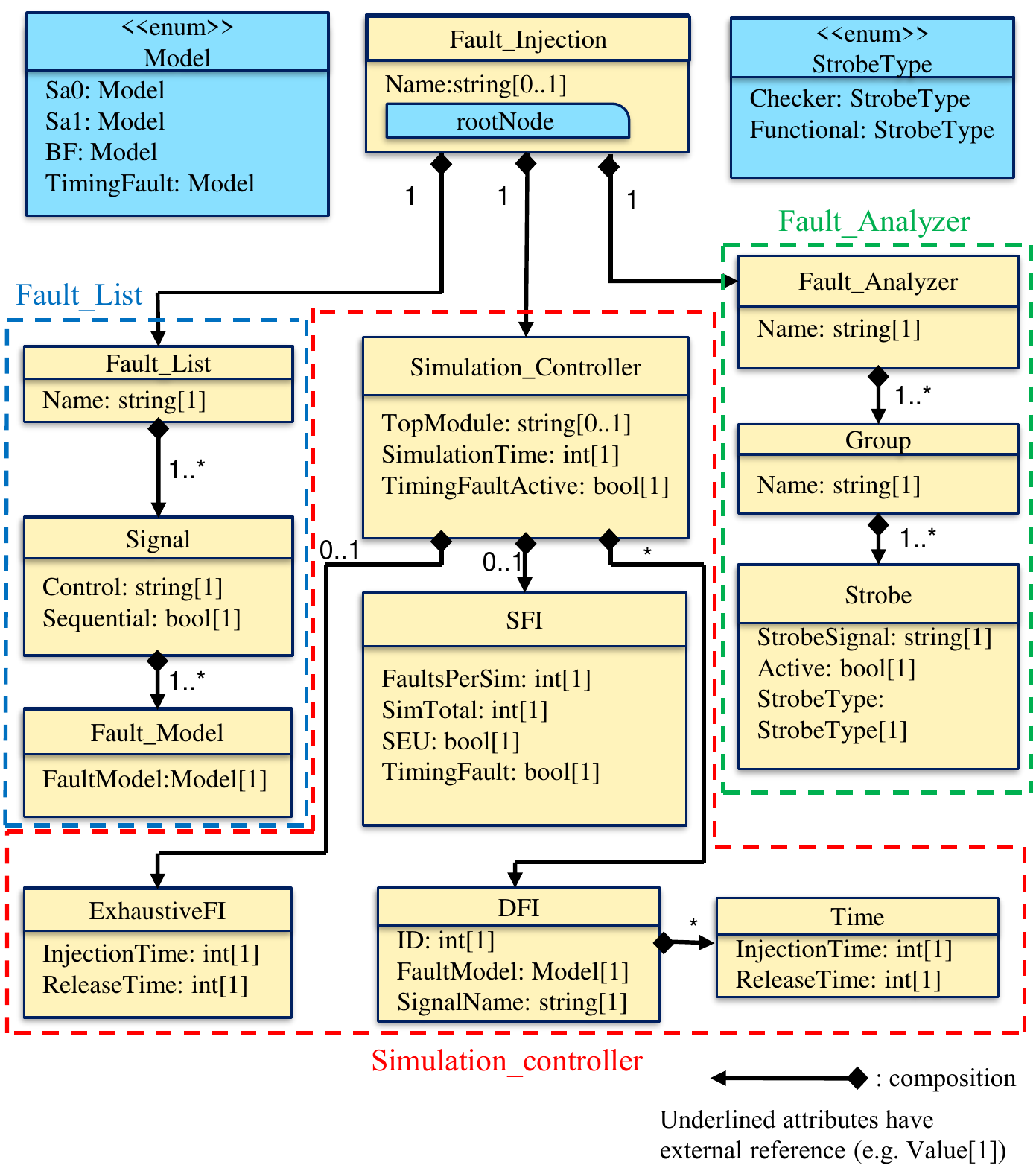}
            		\vspace{-5mm}
            	\end{center}
            	\caption{MetaFI metamodel}
            	\label{fig:MetaFI}
            	\vspace{-5mm}
            \end{figure}

The integral part of the simulation framework is the \textit{MetaFI metamodel}, shown in Fig. \ref{fig:MetaFI}. A model represents a system at a certain abstraction level. Similarly, a metamodel represents the structure of a model and the relation between the elements of a model \cite{Ecker_MM}.
The central class of the metamodel is \textit{Fault\_Injection} that has association relation with three main components of the metamodel:
\begin{itemize}
    \item Simulation\_Controller,
    \item Fault\_List,
    \item Fault\_Analyzer.
\end{itemize}

\textit{Simulation\_Controller} describes the classes that define the fault simulation purpose and fault models. \textit{Simulation\_Controller} class has three attributes: \textit{TopModule}: Top module of the RTL to perform fault simulation, \textit{SimulationTime}: determines for how many clock cycles to run the simulation, \textit{TimingFaultActive}: defines whether timing faults are supported in order to add the extra hardware as presented in section \ref{sec:Extending_the_fault_injection_flow}. This class has a composition relation to three classes: \textit{SFI}, \textit{DFI}, and \textit{ExhaustiveFI}. These three classes narrate the most common fault simulation purposes such as Statistical Fault Injection (SFI), Direct Fault Injection (DFI), and Exhaustive Fault Injection (ExhaustiveFI).

\begin{itemize}
    \item SFI is one of the widely used techniques to determine the dependability of safety-critical designs according to ISO 26262. During an SFI campaign, only a certain random subset of all possible errors is injected. The subset is random in relation to fault injection time and location. This process allows the designer to define the number of experiments to get a rigorous evaluation of the margin of error and confidence interval for the SFI campaign \cite{SFI}. \textit{SFI} class allows the user to select the total number of the injected faults via the attribute \textit{SimTotal}. \textit{FaultPerSim} details the total number of faults to inject per single simulation, thus allowing single and multiple faults per simulation. \textit{SEU} and \textit{TimingFault} are boolean attributes and, when set to \textit{True}, inject only Single Event Upsets or Timing faults respectively. The fault model, fault injection time, and fault location are all randomized.
    
    \item For particular applications, it is needed to analyze the effects of the injected fault on specific locations, specific clock cycle and desired fault model. \textit{DFI} class permits the user to select the certain fault model to inject with the help of \textit{FaultModel} attribute. This attribute has a relation to the \textit{Model} enumeration class that supports four different fault models. \textit{ID} class lists the set of the faults to inject per simulation, i.e. fault signals with the same ID will be injected at the same simulation. \textit{SignalName} class determines the fault location. This class has a relation to \textit{Time} class, specifying the clock cycle to inject and release the fault.
    
    \item Design for Test (DFT) techniques require an exhaustive fault simulation, i.e. to inject all possible stuck-at faults at all possible locations at the desired design component. \textit{ExchaustiveFI} class enables injecting all stuck-at faults at all signals present in the fault list. \textit{InjectionTime} and \textit{ReleaseTime} attribute determine the clock cycle to inject the fault and remove the faulty line, respectively.
\end{itemize}

\textit{Fault\_List} represents the model-based fault list description. A python-based script extracts the information from the design, performs fault collapsing, and populates the model with the necessary fault information. \textit{FaultList} class has a 1..* (one to many) relation to the \textit{Signal} class. It means that each signal of the intended component to apply fault simulation will be added to the fault list. \textit{Control} attribute represents the saboteur control line that manages the fault to inject at the certain signal and \textit{Sequential} attribute determines whether the signal is the input of a register/latch which is important when injecting Single Event Upsets. \textit{Fault\_Model} shows all possible faults to inject at the signal, e.g. after collapsing, only one of two stuck-at fault models is needed.

\textit{Fault\_Analyzer} describes the part of the metamodel that is responsible to collect and analyze the data from the fault injections. \textit{Group} class represents modules and sub-modules of the design, outputs of which are intended to be analyzed. It has a 1..* relation to the \textit{Strobe} class that defines all output signals that should be analyzed. \textit{StrobeSignal} attribute represents the output signal, \textit{Active} defines whether it should be analyzed or not and \textit{StrobeType} determines the output type, whether it is a functional strobe (functional output of the design) or a checker strobe (output of a safety mechanism).

\subsection{Fault simulation algorithm}

\begin{figure}[!tbh]
            \vspace{-4mm}
            	\begin{center}
            		\includegraphics[scale=0.4]{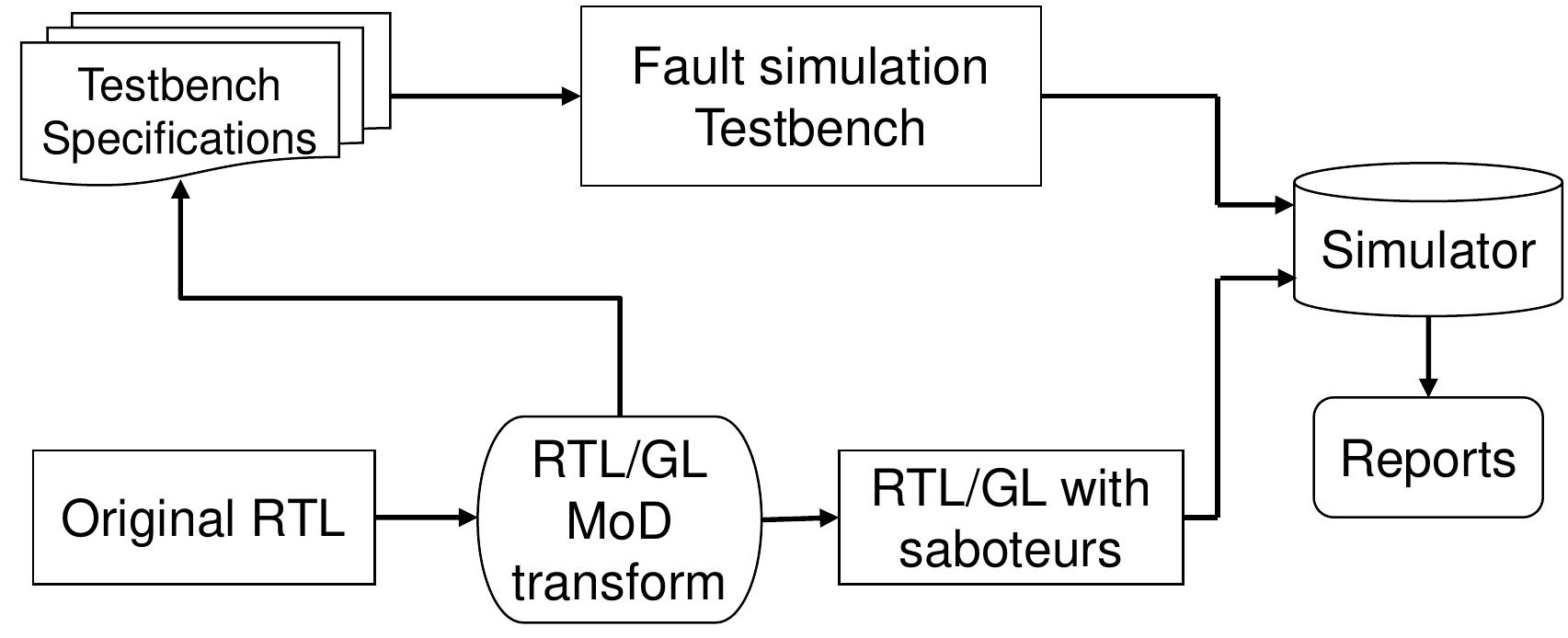}
            		\vspace{-5mm}
            	\end{center}
            	\caption{Fault simulation flow}
            	\label{fig:fault_simulation_flow}
            	\vspace{-4mm}
            \end{figure}

The MetaFI metamodel defines all the necessary features to generate fault simulation testbenches. Using APIs of the internal code generation framework, SystemVerilog (SV), Verilog and C++ fault simulation testbenches are generated conforming to the metamodel classes. The generated testbenches are simulator independent, i.e. all commercial and open-source SV and Verilog simulators can be utilized to perform fault simulation, while C++ generated testbench is specific to Verilator\cite{Verilator}, an open-source widely used simulator in academia. A description of the complete flow is given in Fig. \ref{fig:fault_simulation_flow}.

Fig. \ref{fig:psuedocode} presents the pseudocode of the testbench. Initially, a golden simulation is run without any fault injected. The simulation is run for as many clock cycles as determined by the user (line 4) and the value of the selected strobe signals at every clock cycle is stored in \textit(output\_data) variable as shown in line 6.
Next, the fault simulation procedure is run. \textit{Sim\_count} is the variable counting the number of serial fault simulations, i.e. after the complete simulation is completed, it will increase by one. The testbench checks which fault simulation category is selected by the user (lines 10, 20, 30). If SFI or DFI is selected, a while loop determines how many fault simulations to run as decided by the user (lines 11, 31 respectively.) If ExhaustiveFI is selected, then there should 2 stuck-at fault (s-a-0, s-a-1) simulations for every fault line (line 31). The Design Under Test (DUT) must be in a clean initial state before every simulation is started (lines 12, 22, 32). For each simulation, a set of faults from the fault list is injected complying with the user data (lines 15,26,35), e.g. only stuck-at faults, timing faults, random faults, etc. As the last step, the value of selected strobe lines is compared to the data from the golden simulation. The fault is then classified according to the impact it has on the design and this information is stored in log files.

   \begin{figure}[!tbh]
            \vspace{-2mm}
            \begin{algorithmic}[1]
                \scriptsize
                \Procedure{golden\_sim}{sim\_time}
                \State initial\_state()
                \State fault\_free\_signals()
                \While {time++ $<$ sim\_time}
                \State run\_simulation()
                \State output\_data = store\_data(f\_analyzer.getStrobes())
                \EndWhile
                \State \textbf{return} store\_data()
                \EndProcedure
                
                \Procedure{fault\_sim}{sim\_ctrl, f\_list, f\_analyzer, sim\_time}
                \State sim\_count=0
                \If {sim\_ctrl.hasSFI()}
                    \While{sim\_count $<$ SimTotal}
                    \State  initial\_state()
                    \While {time++ $<$ sim\_time}
                    \State    run\_simulation()
                    \State    inject\_random(f\_list, SEU, TimingFault)
                    \State    output\_data = store\_date(f\_analyzer.getStrobes())   
                    \EndWhile
                    \State compare\_outputs(output\_data,GOLDEN\_SIM(sim\_time))
                    \State sim\_count = sim\_count+1
                    \EndWhile
                \EndIf
                \State sim\_count=0
                \If {sim\_ctrl.hasExhaustiveFI()}
                    \While{sim\_count $<$ total\_fault\_lines}
                    \State  initial\_state()
                    \While {time++ $<$ sim\_time}
                    \State    run\_simulation()
                    \State    inject\_stuck\_at(f\_list)
                    \State    output\_data = store\_date(f\_analyzer.getStrobes())   
                    \EndWhile
                    \State compare\_outputs(output\_data,GOLDEN\_SIM(sim\_time))
                    \State sim\_count = sim\_count+1
                    \EndWhile
                \EndIf
                 \State sim\_count=0
                \If {sim\_ctrl.hasDFI()}
                    \While{sim\_count $<$ length(DFIs)}
                    \State  initial\_state()
                    \While {time++ $<$ sim\_time}
                    \State    run\_simulation()
                    \State    inject\_DFI(f\_list, FaultModel, SignalName, ID)
                    \State    output\_data = store\_date(f\_analyzer.getStrobes())   
                    \EndWhile
                    \State compare\_outputs(output\_data,GOLDEN\_SIM(sim\_time))
                    \State sim\_count = sim\_count+1
                    \EndWhile
                \EndIf
                \EndProcedure
            \end{algorithmic}
            \vspace{-3mm}
            \caption{Fault simulation testbench pseudocode}
            \vspace{-3mm}
            \label{fig:psuedocode}
        \end{figure}

\section{Application and Results}
\label{sec:Application and Results}

The applicability and reduced efforts of fault simulation flow have been demonstrated by running experiments on two different System-on-Chips (SoCs). First SoC is composed of a RISC-V based processor, Instruction and Data Memory, peripherals such as buses, bus bridge, SPI, UART and it is running a \textit{fingerprint calculation application}. This application hashes the value of Program Counter and compares it to an expected value to protect the program control flow.

Second SoC is composed of a RISC-V based processor as well, Instruction and Data Memory, peripherals such as buses, bus bridge, Timer, Interrupt Controller and it is running \textit{Dhrystone application}. Dhrystone measures the processor and compiler performance running a "typical" program.

Different fault simulation strategies were performed on various components of the processor cores. The simulation was run for \textit{10k clock cycles}, i.e. faults were injected only on a random sequence of instructions. Experiments were run on Xcelium \REGISTERED \cite{Xcelium} Verilog simulator.

Tables in the upcoming sub-sections display the results of different fault injection strategies. \textit{Fault set} represents the total number of signals to inject faults and \textit{Total simulations} indicate the total number of fault simulations. \textit{Fail} represents the number of faults that have propagated to the functional strobes, and \textit{Safe} represents silent faults. Further, \textit{Runtime} indicates the complete time it took to run the experiments.

\subsection{Exhaustive Fault Injection}

Table \ref{tab:fingerprint_efi} and Table \ref{tab:dhrystone_efi} show the results of performing exhaustive fault injection into RISC-V CPU components of the SoCs. Both stuck-at-0 and stuck-at-1 faults were injected at all netlist signals. Since fault collapsing is applied to the fault list, the total amount of stuck-at faults is reduced. For example, let us consider ALU in Table \ref{tab:fingerprint_efi}. Fault set contains 2675 signals, i.e. 5350 (2675 times 2) possible stuck-at faults, but after fault collapsing the fault list was reduced to 4713 (12\% reduction). 

\textbf{Observations}: Dhrystone application was more susceptible to stuck-at failures and a notable difference was observed while injecting faults at Execute (EX) stage. Faults injected into the fetch stage (IF) were responsible for the least percentage of failures on both SoCs.

\begin{table}[!tbh]
	\centering
	    \vspace{-5mm}
		\caption{Fingerprint application exhaustive FI}
		\vspace{-3mm}
		\label{tab:fingerprint_efi}
		\begin{tabular}{llllll}
			\hline\hline
			\rule{0ex}{2.5ex} Component & Fault & Total & Fail & Safe & Runtime \\
			                       & set &  simulations& & & h:min \\
			\hline
			\hline
			\rule{0ex}{2.5ex}%
			 ALU  & 2675  & 4713 & 45.0\% & 55.0\% & 02:27 \\
			\hline
			 Prefetcher  & 2546  & 4347 & 55.5\% & 44.5\% & 03:18 \\
			\hline
			 IF stage  & 8449  & 13683 & 20.0\% & 80.0\% & 11:06 \\
			\hline
			EX stage  & 11784  & 20489 & 35.9\% &64.1\%  & 23:24 \\
			\hline
			\hline
		\end{tabular}
		\vspace{-2mm}
	\end{table}
	
\begin{table}[!tbh]
	\centering
	    \vspace{-5mm}
		\caption{Dhrystone application exhaustive FI}
		\vspace{-3mm}
		\label{tab:dhrystone_efi}
		\begin{tabular}{llllll}
			\hline\hline
			\rule{0ex}{2.5ex} Component & Fault & Total & Fail & Safe & Runtime \\
			                       & lines &  simulations& & & h:min \\
			\hline
			\hline
			\rule{0ex}{2.5ex}%
			 ALU  & 2675  & 4713 & 52.1\% & 47.9\% & 02:26 \\
			\hline
			 Prefetcher  & 2546  & 4347 & 71.6\% & 28.4\% & 03:57 \\
			\hline
			 IF stage  & 7948  & 12788 & 32.4\% & 67.6\% & 08:45 \\
			\hline
			EX stage  & 12322  & 21601 & 68.9\% &31.1\%  & 16:21 \\
			\hline
			\hline
		\end{tabular}
		\vspace{-2mm}
	\end{table}

\subsection{Statistical Fault Injection}

Table \ref{tab:fingerprint_sfi} and Table \ref{tab:dhrystone_sfi} show the results of performing SFI on the same components as above. 5000 faults were injected randomly: random clock cycle, random location, and random fault model. Timing faults were not included since the design area would be increased and it would impact the runtime performance.

\textbf{Observations}: Similarly to Exhaustive Fault Injection, Dhrystone application was more susceptible to faults. Nevertheless, we see a reduction of failures on both IF and EX stage and the one reason for that is the small number of simulations compared to the total fault set.

\begin{table}[!tbh]
	\centering
	    \vspace{-5mm}
		\caption{Fingerprint application SFI without timing faults}
		\vspace{-3mm}
		\label{tab:fingerprint_sfi}
		\begin{tabular}{llllll}
			\hline\hline
			\rule{0ex}{2.5ex} Component & Fault & Total & Fail & Safe & Runtime \\
			                       & lines &  simulations& & & h:min \\
			\hline
			\hline
			\rule{0ex}{2.5ex}%
			 ALU  & 2675  & 5000 & 16.4\% & 83.6\%  & 03:10 \\
			\hline
			 Prefetcher  & 2546  & 5000 & 21.2\% & 78.8\% & 03:56 \\
			\hline
			 IF stage  & 8449  & 5000 & 7.9\% & 92.1\% & 04:33 \\
			\hline
			EX stage  & 11784  & 5000 & 10.1\% & 89.9\%  & 05:23 \\
			\hline
			\hline
		\end{tabular}
		\vspace{-2mm}
		\newline
	\end{table}

\begin{table}[!tbh]
	\centering
	    \vspace{-2mm}
		\caption{Dhrystone application SFI without timing faults}
		\vspace{-3mm}
		\label{tab:dhrystone_sfi}
		\begin{tabular}{llllll}
			\hline\hline
			\rule{0ex}{2.5ex} Component & Fault & Total & Fail & Safe & Runtime \\
			                       & lines &  simulations& & & h:min \\
			\hline
			\hline
			\rule{0ex}{2.5ex}%
			 ALU  & 2675  & 5000 & 20.5\% & 79.5\% & 03:15 \\
			\hline
			 Prefetcher  & 2546  & 5000 & 31.6\% & 68.4\% & 03:28 \\
			\hline
			 IF stage  &  7948  & 5000 & 12.6\% & 87.4\% & 04:01 \\
			\hline
			EX stage  & 12322  & 5000 & 11.9\% & 88.1\%  & 04:39 \\
			\hline
			\hline
		\end{tabular}
		\vspace{-2mm}
	\end{table}

\subsection{SEU Fault Injection}

Table \ref{tab:fingerprint_seu} and Table \ref{tab:dhrystone_seu} present the results of randomly injecting only SEU faults into the SoC components. 
\textbf{Observations}: Similarly, Dhrystone application is more susceptible to faults, but as can be seen from the tables, the rate of failures is fairly low for both applications.

\begin{table}[!tbh]
	\centering
	    \vspace{-5mm}
		\caption{Fingerprint application SEU faults}
		\vspace{-3mm}
		\label{tab:fingerprint_seu}
		\begin{tabular}{llllll}
			\hline\hline
			\rule{0ex}{2.5ex} Component & Fault & Total & Fail & Safe & Runtime \\
			                       & lines &  simulations& & & h:min \\
			\hline
			\hline
			 Prefetcher  & 208  & 500 & 14.4\% & 85.6\% & 00:16\\
			\hline
			 IF stage  & 857  & 2000 & 5.3\% & 84.7\% & 01:35 \\
			\hline
			EX stage  & 959  & 2000 & 1.8\%\% & 98.2\% & 01:42  \\
			\hline
			\hline
		\end{tabular}
		\vspace{-1mm}
	\end{table}

\begin{table}[!tbh]
	\centering
	    \vspace{-4mm}
		\caption{Dhrystone application SEU faults}
		\vspace{-3mm}
		\label{tab:dhrystone_seu}
		\begin{tabular}{llllll}
			\hline\hline
			\rule{0ex}{2.5ex} Component & Fault & Total & Fail & Safe & Runtime \\
			                       & lines &  simulations& & & h:min \\
			\hline
			\hline
			 Prefetcher  & 208  & 500 & 18.2\% & 87.8\% & 00:13 \\
			\hline
			 IF stage  & 857  & 2000 & 6.4\% & 83.6\% & 01:12 \\
			\hline
			EX stage  & 926  & 2000 & 3.1\%\% & 96.9\% & 01:32 \\
			\hline
			\hline
		\end{tabular}
		\vspace{-1mm}
	\end{table}

\subsection{Statistical Fault Injection including timing faults}

SFI results presented in Table \ref{tab:fingerprint_sfi} and Table \ref{tab:dhrystone_sfi} are extended to support timing faults as well. As explained in section \ref{sec:Extending_the_fault_injection_flow}, timing faults add an extra register and multiplexer to the saboteur. Table \ref{tab:fingerprint_sfi_timing} and Table \ref{tab:dhrystone_sfi_timing} display the results of conducting SFI campaign including timing faults in the SoCs.

\textbf{Observations}: Timing faults do not increase the rate of failure. Results are very similar to the ones presented in Table \ref{tab:fingerprint_sfi} and Table \ref{tab:dhrystone_sfi} and the marginal difference is only 0-2\%.

\begin{table}[!tbh]
	\centering
	    \vspace{-5mm}
		\caption{Fingerprint application SFI with timing faults}
		\vspace{-3mm}
		\label{tab:fingerprint_sfi_timing}
		\begin{tabular}{llllll}
			\hline\hline
			\rule{0ex}{2.5ex} Component & Fault & Total & Fail & Safe & Runtime \\
			                       & lines &  simulations& & & h:min \\
			\hline
			\hline
			\rule{0ex}{2.5ex}%
			 ALU  & 2675  & 5000 & 15.3\% & 84.7\% & 11:01 \\
			\hline
			 Prefetcher  & 2546  & 5000 & 21.0\% & 79.0\% & 09:14\\
			\hline
			 IF stage  & 8449  & 5000 & 7.5\% & 92.5\% & 25:01 \\
			\hline
			EX stage  & 11784  & 5000 & 10.6\% & 89.4\% & 72:03 \\
			\hline
			\hline
		\end{tabular}
		\vspace{-1mm}
	\end{table}

\begin{table}[!tbh]
	\centering
	    \vspace{-5mm}
		\caption{Dhrystone application SFI with timing faults}
		\vspace{-3mm}
		\label{tab:dhrystone_sfi_timing}
		\begin{tabular}{llllll}
			\hline\hline
			\rule{0ex}{2.5ex} Component & Fault & Total & Fail & Safe & Runtime \\
			                       & lines &  simulations& & & h:min \\
			\hline
			\hline
			\rule{0ex}{2.5ex}%
			 ALU  & 2675  & 5000 & 22.0\% & 78.0\% & 10:05 \\
			\hline
			 Prefetcher  & 2546  & 5000 & 32.2\% & 67.8\% & 09:25 \\
			\hline
			 IF stage  &  7948  & 5000 & 11.6\% & 88.4\% & 29:59 \\
			\hline
			EX stage  & 12322  & 5000 & 12.7\% & 88.1\% & 78:14  \\
			\hline
			\hline
		\end{tabular}
		\vspace{-2mm}
	\end{table}

\subsection{Random timing faults}

As the last experiment, we injected randomly only timing faults into the SoCs and the results are shown on Table \ref{tab:fingerprint_timing} and Table \ref{tab:dhrystone_timing}.

\textbf{Observations}: Similar results as previous SFI campaigns are noticed. Failure rate is higher for Prefetcher injected faults(6\% on average) and the marginal difference for the other components is still fairly low. 

\begin{table}[!tbh]
	\centering
	    \vspace{-5mm}
		\caption{Fingerprint application random timing faults}
		\vspace{-3mm}
		\label{tab:fingerprint_timing}
		\begin{tabular}{llllll}
			\hline\hline
			\rule{0ex}{2.5ex} Component & Fault & Total & Fail & Safe & Runtime \\
			                       & lines &  simulations& & & h:min \\
			\hline
			\hline
			\rule{0ex}{2.5ex}%
			 ALU  & 2675  & 5000 & 15.6\% & 84.4\% & 10:48 \\
			\hline
			 Prefetcher  & 2546  & 5000 & 27.0\% & 77.0\% & 08:50 \\
			\hline
			 IF stage  & 8449  & 5000 & 6.0\% & 94.0\% & 23:11 \\
			\hline
			EX stage  & 11784  & 5000 & 11.5\% & 88.5\% & 70:04  \\
			\hline
			\hline
		\end{tabular}
		\vspace{-1mm}
	\end{table}

\begin{table}[!tbh]
	\centering
	    \vspace{-2mm}
		\caption{Dhrystone application random timing faults}
		\vspace{-3mm}
		\label{tab:dhrystone_timing}
		\begin{tabular}{llllll}
			\hline\hline
			\rule{0ex}{2.5ex} Component & Fault & Total & Fail & Safe & Runtime \\
			                       & lines &  simulations& & & h:min \\
			\hline
			\hline
			\rule{0ex}{2.5ex}%
			 ALU  & 2675  & 5000 & 22.5\% & 77.5\% & 09:34 \\
			\hline
			 Prefetcher  & 2546  & 5000 & 36.2\% & 63.8& 08:43 \\
			\hline
			 IF stage  & 7948  & 5000 & 12.4\% & 87.6\% & 26:50 \\
			\hline
			EX stage  & 12322  & 5000 & 13.0\% & 87.0\% & 76:20 \\
			\hline
			\hline
		\end{tabular}
		\vspace{-1mm}
	\end{table}

\subsection{General observations}
	
The fully automated flow facilitated fault injection campaigns with a very minimal effort of 2 person-days, spent mostly on configuring the fault campaign attributes. Further, the technique was scalable with a full System-on-Chip design, a bottleneck for fault simulators since many are utilized only on a processor level \cite{gemfi}, \cite{Pravadelli}, \cite{Magnusson}. Moreover, \cite{kaja} shows that the existing fault injection flow introduces an overhead of 16\%-99\% of overall runtime compared to the original design, but an increase of 3.5x-8.4x in the fault simulation performance compared to full GL simulation..
After running the experiments, we observed that the IF stage was the component most resilient to failures. This was due to the fact that one of the biggest modules of IF stage is the Exception Unit. Since no exception was detected during the applications run, the fault injected into it had no real effect on the SoC. Noticeably, Prefetcher had the highest rate of failure on every fault simulation campaign, because it is the component that controls the PC.
We noticed also that the failure rate of the applications is low. The main reason is that the random interval (10k clock cycles) of instruction sequences that we chose to inject faults was repeating many series of instructions due to software loops and we expect a higher failure rate for different intervals.
Timing faults introduced an acceptable runtime performance overhead, 3-16.8x slower, considering the increased design area due to added registers.

\section{Related work}
\label{sec:Related work}

Fault simulation is a well-researched topic in academia and industry. In this section, we are providing a summary of the most related works.

Pravadelli et al.\cite{Pravadelli} present an automated non-intrusive simulation-based fault injection framework based on QEMU, an emulator for microprocessor architectures. The authors injected faults into the processor by masking data structures inside QEMU to mimic the behavior of the faults. The framework supports stuck-at faults, timing faults, as well as bit-flips, and experiments were conducted in x86 and ARM processors. This approach provides a fast simulation flow for CPU designs but could lack some accuracy compared to cycle-accurate fault injection and further estimations are needed.

An approach to inject faults on microarchitectural simulators is presented in \cite{genfi}. Authors have extended MARSS and Gem5 simulators to support fault injection (namely MaFIN and GeFIN respectively). A Fault Mask Generator is implemented that can produce a random set of fault masks for different types of faults such as bitflips, permanent and intermittent faults. An Injection Campaign Controller reads the masks and sends injection requests to the Injector Dispatcher, a module that communicates directly with the simulators. Experiments were run on x86 and ARM processors, and the technique is only suitable for processor fault simulation.

In \cite{Kammler} authors propose a framework for Verilog-based fault injection based on Verilog Programming Interface (VPI). A single system function allows the user to determine the fault type, location, and duration of the fault. With the help of callback functions, i.e. using VPI to control the Verilog simulator, the selected fault is injected. The technique can only be applied to Verilog and VPI-compliant simulators.

Rosa et al. \cite{Rosa} present a fault injection framework to evaluate soft error reliability of multi/many-core design. The framework is able to inject faults into the memory area, CPU registers, and interconnection infrastructure. Authors have extended the existing OPVSim and via the help of callbacks, they enabled accessing, modifying, and controlling platform components during simulation, thus, the user can determine the fault injection campaign by setting particular flags to inform target components.

\cite{baraza-etal1}, \cite{baraza-etal2}, \cite{mefisto} modify the VHDL description of the design with the help of saboteurs and mutants. These techniques provide a high degree of controllability but are limited to RTL and VHDL description language.

A fault injection tool, SINJECT, based on the synthesizability of HDL models is proposed in \cite{Zarandi}. The tool provides injection of bitflips and permanent faults into the Verilog and VHDL models. The authors ran experiments on two different small processors, ARP and DP32. Even though the tool provides an automated fault injection campaign, we expect it to have a high overhead when it comes to simulating an SoC and further experiments are needed.

\section{Conclusion and future work}
\label{sec:Conclusion and future work}

We introduced in this paper a fully automated fault simulation framework, namely MetaFI, supporting different permanent and transient fault models. Furthermore, MetaFI enables running automatically various fault injection campaigns such as EFI, SFI, and DFI. Numerous experiments were conducted on different components of two SoCs, running fingerprint calculation application and Dhrystone program respectively. A very minimal manual effort of 2 man-days was required to run 38 distinct fault injection campaigns. Additionally, it was concluded that Prefetcher was the component most susceptible to failures, and protection mechanisms are recommended to be integrated into this component. Timing faults introduced an overhead of 3-16x due to the increased design area.
In future work, we aim to reduce timing faults overhead by dynamically modifying the saboteurs. Also, extending the fault injection flow to use different techniques apart from the simulation is an outlook for future work.

\section{Acknowledgements}
\label{sec:acknowledgements}

Part of the work has been performed in the project ArchitectECA2030 under grant agreement No 877539. The project is co-funded by grants from Germany, Netherlands, Czech Republic, Austria, Norway, France and Electronic Component Systems for European Leadership Joint Undertaking (ECSEL JU).
Part of the work  described  herein  is  also funded  by  the  German Federal Ministry of Education and Research (BMBF) as part of the research project Scale4Edge (16ME0122K). %


\begin{thebibliography}{10}

\bibitem{ISO26262}
{Road vehicles — Functional safety — Part 11: Guidelines on application of
  ISO 26262 to semiconductors}.
\newblock \url{https://www.iso.org/obp/ui/#iso:std:iso:26262:-11:ed-1:v1:en}.

\bibitem{arlat}
J.~Arlat, Y.~Crouzet, and J.-C. Laprie.
\newblock Fault injection for dependability validation of fault-tolerant
  computing systems.
\newblock In {\em [1989] The Nineteenth International Symposium on
  Fault-Tolerant Computing. Digest of Papers}, pages 348--355, 1989.

\bibitem{baraza-etal2}
J.C. Baraza, J.~Gracia, D.~Gil, and P.J. Gil.
\newblock Improvement of fault injection techniques based on vhdl code
  modification.
\newblock In {\em Tenth IEEE International High-Level Design Validation and
  Test Workshop, 2005.}, pages 19--26, 2005.

\bibitem{baraza-etal1}
Juan-Carlos Baraza, JoaquÍn Gracia, Sara Blanc, Daniel Gil, and Pedro-J. Gil.
\newblock Enhancement of fault injection techniques based on the modification
  of vhdl code.
\newblock {\em IEEE Transactions on Very Large Scale Integration (VLSI)
  Systems}, 16(6):693--706, 2008.

\bibitem{Baumann}
R.C. Baumann.
\newblock Radiation-induced soft errors in advanced semiconductor technologies.
\newblock {\em IEEE Transactions on Device and Materials Reliability},
  5(3):305--316, 2005.

\bibitem{FI-techniques}
Alfredo Benso and Paolo Prinetto.
\newblock {\em Fault Injection Techniques and Tools for Embedded Systems
  Reliability Evaluation}.
\newblock 01 2003.

\bibitem{Breuer}
M.A. Breuer.
\newblock The effects of races, delays, and delay faults on test generation.
\newblock {\em IEEE Transactions on Computers}, C-23(10):1078--1092, 1974.

\bibitem{Xcelium}
{Cadence Design Systems}.
\newblock {Xcelium Logic Simulation}.
\newblock
  \url{https://www.cadence.com/ko_KR/home/tools/system-design-and-verification/simulation-and-testbench-verification/xcelium-simulator.html}.

\bibitem{Ecker_MM}
Wolfgang Ecker and Johannes Schreiner.
\newblock {\em Metamodeling and Code Generation in the Springer
  Science+Business Media Dordrecht}, pages 1--41.
\newblock 01 2016.

\bibitem{Pravadelli}
Davide Ferraretto and Graziano Pravadelli.
\newblock Efficient fault injection in qemu.
\newblock In {\em 2015 16th Latin-American Test Symposium (LATS)}, pages 1--6,
  2015.

\bibitem{permanent_fault}
Jörg Henkel, Lars Bauer, Nikil Dutt, Puneet Gupta, Sani Nassif, Muhammad
  Shafique, Mehdi Tahoori, and Norbert Wehn.
\newblock Reliable on-chip systems in the nano-era: Lessons learnt and future
  trends.
\newblock In {\em 2013 50th ACM/EDAC/IEEE Design Automation Conference (DAC)},
  pages 1--10, 2013.

\bibitem{transient_fault}
Andreas Herkersdorf, Michael Engel, Michael Glaß, Jörg Henkel, Veit
  Kleeberger, Michael Kochte, Johannes Kühn, Sani Nassif, Holm Rauchfuss,
  Wolfgang Rosenstiel, Ulf Schlichtmann, Muhammad Shafique, Mehdi Tahoori,
  Jürgen Teich, Norbert Wehn, Christian Weis, and Hans-Joachim Wunderlich.
\newblock Cross-layer dependability modeling and abstraction in system on chip.
\newblock 03 2013.

\bibitem{mefisto}
E.~Jenn, J.~Arlat, M.~Rimen, J.~Ohlsson, and J.~Karlsson.
\newblock Fault injection into vhdl models: the mefisto tool.
\newblock In {\em Proceedings of IEEE 24th International Symposium on Fault-
  Tolerant Computing}, pages 66--75, 1994.

\bibitem{kaja}
Endri Kaja, Nicolas Gerlin, Mounika Vaddeboina, Luis Rivas, Sebastian Prebeck,
  Zhao Han, Keerthikumara Devarajegowda, and Wolfgang Ecker.
\newblock Towards fault simulation at mixed register-transfer/gate-level
  models.
\newblock In {\em 2021 IEEE International Symposium on Defect and Fault
  Tolerance in VLSI and Nanotechnology Systems (DFT)}, pages 1--6, 2021.

\bibitem{genfi}
Manolis Kaliorakis, Sotiris Tselonis, Athanasios Chatzidimitriou, Nikos
  Foutris, and Dimitris Gizopoulos.
\newblock Differential fault injection on microarchitectural simulators.
\newblock In {\em 2015 IEEE International Symposium on Workload
  Characterization}, pages 172--182, 2015.

\bibitem{Kammler}
David Kammler, Junqing Guan, Gerd Ascheid, Rainer Leupers, and Heinrich Meyr.
\newblock A fast and flexible platform for fault injection and evaluation in
  verilog-based simulations.
\newblock In {\em 2009 Third IEEE International Conference on Secure Software
  Integration and Reliability Improvement}, pages 309--314, 2009.

\bibitem{Kochte}
Michael~A. Kochte, Christian~G. Zoellin, Rafal Baranowski, Michael~E. Imhof,
  Hans-Joachim Wunderlich, Nadereh Hatami, Stefano~Di Carlo, and Paolo
  Prinetto.
\newblock Efficient simulation of structural faults for the reliability
  evaluation at system-level.
\newblock In {\em 2010 19th IEEE Asian Test Symposium}, pages 3--8, 2010.

\bibitem{syfi}
Dongwoo Lee and Jongwhoa Na.
\newblock A novel simulation fault injection method for dependability analysis.
\newblock {\em IEEE Design Test of Computers}, 26(6):50--61, 2009.

\bibitem{SFI}
R.~Leveugle, A.~Calvez, P.~Maistri, and P.~Vanhauwaert.
\newblock Statistical fault injection: Quantified error and confidence.
\newblock In {\em 2009 Design, Automation Test in Europe Conference
  Exhibition}, pages 502--506, 2009.

\bibitem{F-collapsing}
A.~Lloy.
\newblock Advanced fault collapsing (logic circuits testing).
\newblock {\em IEEE Design Test of Computers}, 9(1):64--71, 1992.

\bibitem{Luong}
G.M. Luong and D.M.H. Walker.
\newblock Test generation for global delay faults.
\newblock In {\em Proceedings International Test Conference 1996. Test and
  Design Validity}, pages 433--442, 1996.

\bibitem{Magnusson}
P.S. Magnusson, M.~Christensson, J.~Eskilson, D.~Forsgren, G.~Hallberg,
  J.~Hogberg, F.~Larsson, A.~Moestedt, and B.~Werner.
\newblock Simics: A full system simulation platform.
\newblock {\em Computer}, 35(2):50--58, 2002.

\bibitem{Malaiya}
Yashwant~K. Malaiya and Ramesh Narayanaswamy.
\newblock Modeling and testing for timing faults in synchronous sequential
  circuits.
\newblock {\em IEEE Design Test of Computers}, 1(4):62--74, 1984.

\bibitem{Oetjens}
J.-H. Oetjens, N.~Bannow, M.~Becker, O.~Bringmann, A.~Burger, M.~Chaari,
  S.~Chakraborty, R.~Drechsler, W.~Ecker, K.~Grüttner, Th. Kruse, C.~Kuznik,
  H.~M. Le, A.~Mauderer, W.~Müller, D.~Müller-Gritschneder, F.~Poppen,
  H.~Post, S.~Reiter, W.~Rosenstiel, S.~Roth, U.~Schlichtmann, A.~von Schwerin,
  B.-A. Tabacaru, and A.~Viehl.
\newblock Safety evaluation of automotive electronics using virtual prototypes:
  State of the art and research challenges.
\newblock In {\em 2014 51st ACM/EDAC/IEEE Design Automation Conference (DAC)},
  pages 1--6, 2014.

\bibitem{gemfi}
Konstantinos Parasyris, Georgios Tziantzoulis, Christos~D. Antonopoulos, and
  Nikolaos Bellas.
\newblock Gemfi: A fault injection tool for studying the behavior of
  applications on unreliable substrates.
\newblock In {\em 2014 44th Annual IEEE/IFIP International Conference on
  Dependable Systems and Networks}, pages 622--629, 2014.

\bibitem{Rosa}
Felipe Rosa, Fernanda Kastensmidt, Ricardo Reis, and Luciano Ost.
\newblock A fast and scalable fault injection framework to evaluate
  multi/many-core soft error reliability.
\newblock In {\em 2015 IEEE International Symposium on Defect and Fault
  Tolerance in VLSI and Nanotechnology Systems (DFTS)}, pages 211--214, 2015.

\bibitem{Schreiner.etal-2016}
J.~Schreiner, R.~Findenig, and W.~Ecker.
\newblock {Design Centric Modeling of Digital Hardware}.
\newblock In {\em {IEEE} International High Level Design Validation and Test
  Workshop, {HLDVT} 2016}, pages 46--52, 2016.

\bibitem{Schreiner.etal-2017}
J.~Schreiner, F.~Willgerodt, and W.~Ecker.
\newblock A new approach for generating view generators.
\newblock In {\em Design and Verification Conference - US}, Feb 2017.
\newblock \url{http://events.dvcon.org/2017/proceedings/papers/04P_18.pdf}.

\bibitem{Verilator}
{Verilator}.
\newblock {Verilator}.
\newblock \url{https://www.veripool.org/verilator/}.

\bibitem{Zarandi}
H.R. Zarandi, S.G. Miremadi, and A.~Ejlali.
\newblock Dependability analysis using a fault injection tool based on
  synthesizability of hdl models.
\newblock In {\em Proceedings 18th IEEE Symposium on Defect and Fault Tolerance
  in VLSI Systems}, pages 485--492, 2003.

\bibitem{ziade}
Haissam Ziade, Rafic Ayoubi, and R.~Velazco.
\newblock A survey on fault injection techniques.
\newblock {\em Int. Arab J. Inf. Technol.}, 1:171--186, 01 2004.

\end{thebibliography}

\end{document}